\documentclass[twocolumn,showpacs,showkeys,superscriptaddress]{revtex4}
\usepackage{amsmath}
\usepackage{amsfonts}
\usepackage{amssymb}
\usepackage{color}

\begin{document}

\title{Generalized Magnetofluid Connections in Pair Plasmas}
\author{Felipe A. Asenjo}
\email{felipe.asenjo@uai.cl}
\affiliation{Facultad de Ingenier\'{\i}a y Ciencias, Universidad Adolfo Ib\'a\~nez, Santiago 7941169, Chile.}
\author{Luca Comisso}
\email{luca.comisso@polito.it}
\affiliation{Department of Astrophysical Sciences and Princeton Plasma Physics Laboratory, Princeton University, Princeton, New Jersey 08543, USA.}
\affiliation{Dipartimento di Energia, Politecnico di Torino, 10129, Torino, Italy and Istituto dei Sistemi Complessi - CNR, 00185, Roma, Italy.}
\author{Swadesh M. Mahajan}
\email{mahajan@mail.utexas.edu}
\affiliation{Institute for Fusion Studies, The University of Texas at Austin, Texas 78712, USA.}

\date{\today}

\begin{abstract}
We extend the magnetic connection theorem of ideal magnetohydrodynamics to nonideal relativistic pair plasmas. Adopting a generalized Ohm's law, we prove the existence of generalized magnetofluid connections that are preserved by the plasma dynamics. We show that these connections are related to a general antisymmetric tensor that unifies the electromagnetic and fluid fields. The generalized magnetofluid connections set important constraints on the plasma dynamics by forbidding transitions between configurations with different magnetofluid connectivity. An approximated solution is explicitly shown  where the corrections due to current inertial effects are found.
\end{abstract}


\maketitle

One of the most important ideas in plasma physics is that in an ideal magnetohydrodynamical plasma, two plasma elements connected by a magnetic field line at a given time will remain connected by a magnetic field line at any subsequent time. As it has been shown in a classic paper by Newcomb \cite{Newcomb}, this occurs because a plasma that satisfies the ideal Ohm's law moves with a transport velocity that preserves the ``magnetic connections" between plasma elements. 

The implications of this ``magnetic connection theorem" are profound and allow us to give a meaning to fundamental concepts in classical plasma physics, such as magnetic field line motion, magnetic topology, and magnetic reconnection. However, these concepts have to be taken with care when considering plasmas in the relativistic regime. Indeed, in relativisic plasmas other difficulties arise, as the distinction between the magnetic field and the electric field, which are not independent from the reference frame. These issues have been considered by Pegoraro \cite{pegoraroEPJ}, who has shown that the magnetic connection theorem of ideal magnetohydrodynamics (MHD) can be cast in a covariant form, but its interpretation in terms of magnetic field lines alone requires a time resetting of the connected plasma elements in such a way to restore simultaneity when the reference frame is changed.

Recently, a more general connection theorem has been proved for nonideal electron-ion plasmas \cite{asenjoComissoCon}, showing the existence of ``generalized magnetofluid connections" that are preserved during the evolution of a dissipationless system. In the nonideal case the magnetic connections are replaced by more general connections that are linked to an antisymmetric tensor that unifies the electromagnetic and fluid fields \cite{asenjoComissoCon}. This has important consequences on the plasma dynamics, which is constrained by the presence of the generalized magnetofluid connections.

The general theory of unifying the electromagnetic and fluid fields (fully relativistic) was developed, in the modern context, by Mahajan \cite{mahajanU} though an earlier work of Bekenstein \cite{Bekenstein_1987}, had broached the subject. The essence of this encompassing formalism is the construction of a fully antisymmetric Generalized Vorticity  tensor ${\cal M}^{\mu\nu}$ whose spatial components define the generalized vorticity vector. This tensor also serves as a basis for demonstrating that, for covariant  ideal dynamics, there exists a helicity invariant that constraints the motions accessible to an ideal fluid. The basic concepts introduced in \cite{mahajanU} have been 
 been successfully  extended and applied to diverse plasma frameworks  \cite{Bambah_2006, MahajanAsenjoPRL2011, mahajan2010,mahajan2011,asenjoGR,Gao_2014,Bhattacharjee_2015,Yoshida_2014,Yoshida2_2014}; additional physics adds further complexity to ${\cal M}^{\mu\nu}$. 

The purpose of this paper is to extend the generalized magnetofluid connection theorem to pair plasmas, i.e., plasmas consisting of negatively and positively charged particles with equal mass and absolute charge. 
Pair plasmas such as electron-positron plasmas are thought to exist abundantly in the early universe \cite{Gibbon_1983,Weinberg_2008} and in many astrophysical environments \cite{Begelman_1984,Michel_1991,Piran_2004}. 
Accordingly, a detailed understanding of the pair plasma dynamics is of crucial importance, and indeed it is currently a subject of extensive study \cite{ComissoAsenjo2014,Helander2014,Kagan2015,Liu2015,Sarri_2015,Lopez2015,Asenjoa,Asenjob,Asenjoc}.

In the following, we consider a Minkowskian spacetime ${x^\mu } = \left( {{x^0},{x^1},{x^2},{x^3}} \right)$
where the line element $ds$ is given by $d{s^2} =  - {(d{x^0})^2} + {(d{x^1})^2} + {(d{x^2})^2} + {(d{x^3})^2} = {\eta _{\mu \nu }}d{x^\mu }d{x^\nu }$.
Also, we choose units in which the speed of light $c$ is unity.
The magnetohydrodynamical pair plasma is specified by a proper particle number density $n$, proper charge density $q=ne$, enthalpy density $h$, four-velocity $U^\mu$ (with the normalization $U_\mu U^\mu=-1$), and normalized four-current density $J^\mu$. The four-velocity can be expressed as $U^\mu=(\gamma,\gamma \vec v)$, with $\gamma = {(1-v^2)^{-1/2}}$ indicating the relativistic kinematic factor.
We recall that for the relativistic perfect fluid in thermal equilibrium \cite{Chandra1938,Synge1957} the enthalpy becomes $h=mnf$, where $f=K_3(\zeta)/K_2(\zeta)$ is the relativistic thermal factor, with $K_n$ indicating the modified Bessel function of the second kind of order $n$ and $\zeta=m/k_B T$ defining the ratio of rest energy to thermal energy.

For a pair plasma, the equations that govern the dynamics  can be deduced from a two-fluid model \cite{koide}. They are the continuity equation 
\begin{equation}\label{continuityeq}
\partial_\mu (qU^\mu)=0\, ,
\end{equation}
the generalized momentum equation
\begin{equation}\label{momentumEquation}
\partial_\nu\left(hU^\mu U^\nu+\frac{h}{4q^2}J^\mu J^\nu\right)=-\partial^\mu p+J_\nu F^{\mu\nu}\, ,
\end{equation}
and the generalized Ohm's law
\begin{eqnarray}\label{ohmlaw1orig}
\frac{1}{4q}\partial_\nu\left[\frac{h}{q}(U^\mu J^\nu+J^\mu U^\nu)\right]=-\frac{1}{2q}\partial^\mu \Delta p+U_\nu F^{\mu\nu}+R^\mu\, .
\end{eqnarray}
In Eq. \eqref{momentumEquation}, the current inertia effects are shown in the left-hand side. Also, in  Eq. \eqref{ohmlaw1orig},   the inertia effect of the current is retained in the left-hand side, while the thermal electromotive effects appear in the first term on the right-hand side.
Besides, $\Delta p$ denotes the proper pressure difference between the positively and negatively charged fluids, while 
\begin{equation}\label{}
  R^\mu= - \eta\left[J^\mu+Q(1+\Theta)U^\mu\right]\, 
\end{equation}
is the frictional four-force density between the fluids, and $Q=U_\alpha J^\alpha$. The resistivity can be recognized as $\eta$, while $\Theta$ represents the thermal energy exchange rate from the negatively to the positively charged fluid (a closed expression of $\Theta$ for pair plasmas is given in Ref.~\cite{koide}).

 $F^{\mu \nu} = {\partial^\mu}{A^\nu} - {\partial^\nu}{A^\mu}$ is the electromagnetic field tensor ($A^\mu$ is the four-vector potential), which obeys Maxwell equations
\begin{equation}\label{}
  \partial_\nu F^{\mu\nu}=4\pi J^\mu\, ,\qquad
 \partial_\nu F^{*\mu\nu}=0\, .
\end{equation}
$F^{*\mu\nu} = (1/2) \epsilon^{\mu\nu\alpha\beta} F_{\alpha\beta}$ is the dual of $F^{\mu\nu}$, and $\epsilon^{\mu\nu\alpha\beta}$ indicates the Levi-Civita symbol. Notice that $\partial_\nu J^\nu=0$.

It was recently been shown in Ref.~\cite{asenjoComissoCon} that,  although different relativistic and nonideal effects could complicate considerably the governing equations of the plasma, the insight to obtain a general connection theorem is to cast the relevant Ohm's law in an antisymmetric form. This allows us to prove the existence of preserved connections when dissipation is negligible.
This is done through the introduction of an antisymmetric tensor that unifies the  electromagnetic and fluid fields. 
As a starting point, we can re-write the generalized Ohm's law \eqref{ohmlaw1orig} in the form
\begin{equation}\label{ohmlaw2}
 {\Sigma}^\mu+\partial^\mu\left(\frac{hQ}{2q^2}\right)={ U}_\nu {\cal M}^{\mu\nu}+ {R}^\mu\, ,
\end{equation}
where we have defined the antisymmetric tensor
\begin{eqnarray}\label{magnetofluidUnifiedTensor}
  {{\cal M}^{\mu \nu}}={F}^{\mu\nu}+\Lambda^{\mu\nu}\, ,
\end{eqnarray}
with the antisymmetric current-field tensor
\begin{eqnarray}\label{}
  \Lambda^{\mu\nu}= \partial^\mu\left(\frac{h}{4q^2}{J}^\nu\right)-\partial^\nu\left(\frac{h}{4q^2}{J}^\mu\right) \, .
\end{eqnarray}
The tensor $\Sigma^\mu$ in \eqref{ohmlaw2} is defined as
\begin{equation}\label{sigmaDp}
\Sigma^\mu=\frac{1}{2q}\partial^\mu\Delta p-\frac{h}{4q}\partial^\mu\left(\frac{Q}{q}\right)-\frac{1}{4q}{J}_\nu S^{\mu\nu}\, ,
\end{equation}
where the antisymmetric flow-field tensor is defined as
\begin{eqnarray}\label{fluidVortic}
  S^{\mu\nu}= \partial^\mu\left(\frac{h}{q}U^\nu\right)-\partial^\nu\left(\frac{h}{q}U^\mu\right) \, .
\end{eqnarray}

The tensor ${\cal M}^{\mu\nu}$, given in Eq.~\eqref{magnetofluidUnifiedTensor}, represents an effective field tensor that unifies the electromagnetic and current-fluid forces. It turns out to be crucial in revealing fundamental properties of the system. This tensor is similar in nature (but different in structure) to the unified magnetofluid field tensors introduced for one-species \cite{mahajanU} and electron-ion \cite{asenjoComissoCon} relativistic plasmas. On the other hand, we can see that the tensor $\Sigma^{\mu}$ of Eq.~\eqref{sigmaDp} contains the fluid-current interplay of the pair plasma, as well as the thermal effects.

To find the evolution of the generalized magnetofluid tensor, we have to take  the curl of Eq.~\eqref{ohmlaw2}. Applying first $\epsilon_{\alpha\beta\gamma\mu}\partial^\gamma$ and then $\epsilon^{\alpha\beta\lambda\phi}$, we obtain
\begin{eqnarray}\label{Important}
  \frac{d {\cal M}^{\lambda\phi}}{d\tau}&=&\partial^\lambda {U}_\nu {\cal M}^{\phi\nu}-\partial^\phi { U}_\nu {\cal M}^{\lambda\nu}\nonumber\\
&&-\partial^\lambda {\Sigma}^\phi+\partial^\phi {\Sigma}^\lambda +\partial^\lambda {R}^\phi-\partial^\phi {R}^\lambda \, ,
\end{eqnarray}
where $d/d\tau\equiv U_\mu\partial^\mu$ is the convective derivative.
Eq.~\eqref{Important} is the antisymmetric form of the generalized Ohm's law for pair plasmas.

Now we can demonstrate that when dissipation is negligible (when the evolution of the system is fast compared to the dissipation time scale), there exist generalized magnetofluid connections that are preserved during the dynamics of the pair plasma. 
Similarly to previous approaches \cite{Newcomb,pegoraroEPJ,asenjoComissoCon}, let us define a displacement four-vector $\Delta x^\mu$ of a plasma element that is transported by the four-velocity 
\begin{equation}\label{dX}
  \frac{\Delta x^\mu}{\Delta\tau} = {U}^\mu+{D}^\mu\, ,
\end{equation}
where $\Delta\tau$ is the variation of the proper time and ${D}^\mu$ is a four-vector field which satisfies the equation \cite{asenjoComissoCon}
\begin{equation}\label{eqD}
 \partial^\mu {D}_\nu={\cal N}_{\alpha\nu}\partial^\mu\Sigma^\alpha\, ,
\end{equation}
with  ${\cal N}_{\mu\nu}$ as the inverse of the ${\cal M}_{\mu\nu}$ matrix (${\cal M}^{\mu\alpha}{\cal N}_{\alpha\nu}={\delta^\mu}_\nu$).
The four-vector ${D}^\mu$ contains the inertial-thermal-current information of ${\Sigma}^{\mu}$. Now, let us also introduce the space-like event-separation four-vector $d l^\mu$ of an element (such that two events are simultaneous in a frame where $dl^0=0$) \cite{pegoraroEPJ}.
Thus, this four-vector is transported by the four-velocity 
\begin{equation}\label{dldt}
  \frac{d }{d\tau}dl^\mu=dl^\alpha\partial_\alpha \left({U}^\mu+D^\mu\right)\, .
\end{equation}

Using Eqs.~\eqref{Important} and \eqref{dldt}, and neglecting the frictional four-force density $R^\mu$, we can prove that
\begin{eqnarray}
  \frac{d }{d\tau}\left(dl_\lambda {\cal M}^{\lambda\phi}\right)
   &=&-\partial^\phi { U}_\nu \left(dl_\lambda {{\cal M}}^{\lambda\nu}\right)+dl_\alpha\partial^\alpha {D}_\lambda {\cal M}^{\lambda\phi}\nonumber\\
&&-dl_\alpha\left(\partial^\alpha\Sigma^\phi-\partial^\phi\Sigma^\alpha\right)\, .
\end{eqnarray}
With the help of Eq.~\eqref{eqD}, the previous equation finally is written as
\begin{equation}\label{CTequation}
  \frac{d }{d\tau}\left(dl_\lambda {\cal M}^{\lambda\phi}\right) =-\left(dl_\lambda {{\cal M}}^{\lambda\nu}\right)\partial^\phi\left({U}_\nu+D_\nu \right)\, .
\end{equation} 
From this equation follows that if $dl_\lambda {\cal M}^{\lambda\phi}=0$ initially, then it will remain null at all times (regularity properties of the four-velocity field \eqref{dX} are assumed). This implies the existence of generalized magnetofluid connections that are preserved during the nonideal pair plasma dynamics. Thereby, the ``magnetofluid connection equation'' \eqref{CTequation} generalizes the standard magnetic connection concept \cite{Newcomb} for pair plasmas with the inclusion of relativistic, thermal-inertial, thermal electromotive, and current-inertia effects. In particular, given that
\begin{equation}\label{generalizedmagneticconnectionsM}
d{l_\lambda }{{\cal M}^{\lambda \phi }} = d{l_\lambda }{F^{\lambda \phi }}+d{l_\lambda }{{\Lambda}^{\lambda \phi }} \, ,
\end{equation}
the generalized magnetofluid connections reduce to the well-known magnetic connections if $d{l_\lambda }{\Lambda^{\lambda \phi }} \to 0$ and $dl_0=0$. In the more general case, in the frame where $dl_0=0$, the condition $d{l_\lambda }{{\cal M}^{\lambda \phi }} = 0$ can be written as the vectorial conditions
\begin{equation}\label{}
d\vec l\times \vec {\cal B}=0 \, , \quad  d\vec l\cdot \vec {\cal E}=0 \, ,
\end{equation}
where 
\begin{equation}\label{}
{\cal B}^k = \frac{1}{2} \epsilon^{ijk} {\cal M}_{ij} \, , \quad  {\cal E}^i = {\cal M}^{0i} \, ,
\end{equation}
are the components of a generalized magnetic-like field and a generalized electric-like field, respectively. On the other hand, if $dl_0 \ne 0$, the simultaneity can be recovered resetting the time \cite{pegoraroEPJ} by changing $dl_\mu\rightarrow d{l'}_\mu=dl_\mu+({U}_\mu+D_\mu) d\lambda$,
such that in this new reference frame $dl'_0=0$. 

It is important to emphasize that the matrix ${\cal N}_{\mu\nu}$ exists since ${\cal M}_{\mu\nu}$ is non-sigular for non-ideal MHD \cite{asenjoComissoCon}, as the determinant of ${\cal M}_{\mu\nu}$ is $\left \| {\cal M} \right\| =(\vec{\cal E}\cdot\vec{\cal B})^2\neq 0$ in general. We can see this from the spatial component of Eq.~\eqref{ohmlaw2}  ${\cal U}^0\vec{\cal E}+\vec{\cal U}\times\vec{\cal B}=\vec{\Sigma}+\nabla(hQ/2q^2)-\vec{R}$, implying that ${\cal U}^0 \vec{\cal E}\cdot \vec{\cal B}=\left(\vec\Sigma+\nabla(hQ/2q^2)-\vec R\right)\cdot\vec{\cal B}$.
Even in the case of vanishing resistivity, it could be $\vec{\cal E}\cdot\vec{\cal B} \neq 0$ due to the non-ideal effects in $\vec{\Sigma}$.

Along with the previous description, as an example, we can find an approximated solution for the field $D^\mu$ given by Eq.~\eqref{eqD} for quasi-uniform electromagnetic fields. 
 Neglecting thermal-inertial effects, as well as variations of $Q/q$, we can approximate
\begin{equation}\label{apprxSigma}
\Sigma^\mu\approx -\frac{1}{4q}{J}_\nu S^{\mu\nu}\, .
\end{equation}
Now we need to make use of the momentum equation \eqref{momentumEquation} written in the general form
\begin{equation}\label{momentumEquation2}
U_\nu S^{\nu\mu}=\frac{1}{q}{J}_\nu {\cal M}^{\mu\nu}+H^\mu\, ,
\end{equation}
with
\begin{equation}
H^\mu=\left(1-\frac{1}{4q^2}{J}_\alpha{J}^\alpha\right)\partial^\mu\left(\frac{h}{q}\right)-\frac{h}{8q}\partial^\mu\left(\frac{1}{q^2}{J}_\alpha{J}^\alpha  \right)-\frac{1}{q}\partial^\mu p\, .
\end{equation}

Considering the above approximations, altogheter with ${J}_\alpha{J}^\alpha\ll q^2$, we get that $U_\nu S^{\nu\mu}\approx (1/q){J}_\nu {\cal M}^{\mu\nu}$, which gives the constraint ${J}_\nu {\cal M}^{\mu\nu}U_\mu=0$. Then, from Eq.~\eqref{momentumEquation2} we can get the approximated solution
\begin{equation}\label{momentumEquation3}
S^{\nu\mu}\approx-\frac{1}{q}{J}_\alpha {\cal M}^{\mu\alpha}U^\nu+\frac{1}{q}{J}_\alpha {\cal M}^{\nu\alpha}U^\mu \, .
\end{equation}
 Using this approximated solution in \eqref{apprxSigma} we find that
\begin{equation}\label{apprxSigma2}
\Sigma^\mu\approx -\frac{Q}{4q^2}{J}_\nu {\cal M}^{\mu\nu}\, .
\end{equation}
In this way,  Eq.~\eqref{eqD} becomes
\begin{equation}\label{eqD2}
 \partial^\mu {D}_\nu\approx{\cal N}_{\nu\alpha}\partial^\mu\left(\frac{Q}{4q^2}{J}_\lambda {\cal M}^{\alpha\lambda} \right)\, ,
\end{equation}
that has a simple approximated solution for quasi-uniform electromagnetic fields \cite{asenjoComissoCon}
\begin{equation}\label{eqD3}
{D}_\nu\approx\frac{Q}{4q^2}{J}_\nu \, ,
\end{equation}
indicating that, under these approximations, the  magnetofluid connections are transported by the four-velocity
\begin{equation}
U^\mu+\frac{Q}{4q^2}{J}^\mu=U_\nu\left({\eta}^{\nu\mu}+\frac{1}{4q^2}J^\nu J^\mu\right)\, ,
\end{equation}
that contains the corrections due to the current inertia effects.

The above generalization of the magnetic connection theorem of ideal MHD to nonideal relativistic pair plasmas gives us an important theoretical framework for investigating high-energy pair plasmas. In this extended framework, the generalized magnetofluid field tensor ${\cal M}^{\mu \nu}$ plays the role that the electromagnetic field tensor $F^{\mu \nu}$ has in the ideal relativistic MHD description. Therefore, when dissipationless nonideal effects are considered, the preserved connections are no longer related to ${F^{\mu \nu}}$ but to ${{\cal M}^{\mu \nu}}$. In particular, if there is simultaneity between two events, the connections given by the field lines of $\vec B$ are replaced by those given by the field lines of $\vec {\cal B}$.
Moreover, we have shown that the plasma elements moves with the general velocity \eqref{dldt}. For a general solution of this velocity, the complete Eq.~\eqref{eqD} must be solved.

On the other hand, as well as in the ion-electron case  developed in Ref.~\cite{asenjoComissoCon}, the magnetic  connection concept for pair plasma is well defined as $U_\nu$ be time-like.  For example, in the ideal case the magnetic connection concept holds when $|\vec{E}| < |\vec{B}|$,  as this condition defines a time-like velocity of the plasma elements.  For the non-ideal pair plasma there is no a simple condition. The dynamical equations \eqref{continuityeq}, \eqref{momentumEquation} and \eqref{ohmlaw1orig} require the time-like condition $U_\mu U^\mu=-1$ for the four-velocity \cite{koide}. This is translated, with the help of  the Ohm's law \eqref{ohmlaw2}, to the condition
\begin{equation}
{\cal N}^{\alpha\beta} \left[\Sigma_\beta+\partial_\beta\left( \frac{hQ}{2q^2}\right)\right] \left[\Sigma^\mu+\partial^\mu\left( \frac{hQ}{2q^2}\right)\right]{\cal N}_{\mu\alpha}=1\, ,
\end{equation}
over the thermal-inertial and fluid-current non-ideal effects (when the frictional four-force density is neglected).

The generalized magnetofluid connections set important constraints on the pair plasma dynamics by forbidding transitions between configurations with different magnetofluid connectivity. This is an essential tool for the understanding of formation of small scale structures occurring in this kind of plasmas plasmas. 
The small scale structures formation (and their specific "shape") is a consequence of the constraint imposed by Eq.~\eqref{CTequation}, where their "shape" is dictated by the advective velocity defined by $U_\nu + D_\nu$. 
Hence, this work  contributes to the understanding of the non-ideal processes in pair plasmas which can be useful for the study of plasmas in  astrophysical environments. Also, it can used to verify the results of the numerical simulations of pair plasmas.

{\it Acknowledgments.} FAA thanks to Fondecyt-Chile for Funding No. 11140025.

\end{document}